\documentclass[twocolumn,preprintnumbers,amsmath,amssymb]{revtex4}
\usepackage{graphicx}
\usepackage{epsfig}
\usepackage{amsmath}
\usepackage{amsfonts}
\usepackage{amssymb}
\usepackage{textcomp}
\usepackage{dcolumn}
\usepackage{feynmf}

\begin{document}

\title{Relation of a Unified Quantum Field Theory of Spinors to the Structure of\\ General Relativity}

\author{Martin Kober}

\email{kober@th.physik.uni-frankfurt.de}

\affiliation{Institut f\"ur Theoretische Physik, Johann Wolfgang Goethe-Universit\"at, Max-von-Laue-Str.~1, 60438 Frankfurt am Main, Germany}

\begin{abstract}
Based on a unified quantum field theory of spinors assumed to describe all matter fields and their interactions we construct the space time structure of general relativity according to a general connection within the corresponding spinor space. The tetrad field and the corresponding metric field are composed from a space time dependent basis of spinors within the internal space of the fundamental matter field. Similar to twistor theory the Minkowski signature of the space time metric is related to this spinor nature of elementary matter, if we assume the spinor space to be endowed with a symplectic structure. The equivalence principle and the property of background independence arise from the fact that all elementary fields are composed from the fundamental spinor field. This means that the structure of space time according to general relativity seems to be a consequence of a fundamental theory of matter fields and not a presupposition as in the usual setting of relativistic quantum field theories.
\end{abstract}


\maketitle

\section{Introduction}

One of the central problems in contemporary theoretical physics is the unification of quantum theory with general relativity.
This problem becomes manifest with respect to the question of the relation between relativistic quantum field theories being formulated on a given Minkowski space time background and yielding the conceptual framework of the standard model of particle physics and the general covariant description of space time and gravity according to general relativity.
In this paper there is made the attempt to derive the properties of the structure of space time according to general relativity from a fundamental quantum field theory of spinors. Such a unified quantum field theory of spinors with a self interaction term of the fundamental spinor field being the origin of mass and interaction of particles has been suggested by Heisenberg. We will start from this theory but formulate it in a setting being background independent in a rigorous sense.
In Heisenberg's original setting the theory was formulated on a given Minkowski background. In contrast to Heisenberg we will not presuppose an a priori metric structure of space time, not even in the sense of general relativity.

The paper is structured as following. It consists of three parts. In the first part there is first presented the basic idea of the unified quantum field theory of spinors according to Heisenberg. After this it is given a short introduction to a description of the space time metric in terms of twistors in the sense developed by Penrose.
With the help of these ideas introduced in the first part in the second part there is formulated a theory where the metric structure of space time appears as a consequence of a symplectic structure combined with a general connection within the abstract space of the fundamental spinor field. The action for the fundamental spinor field is formulated by using a metric constructed from a basis of spinor fields corresponding to this connection from which there is also built an action for gravity. In the third section there is made an attempt to perform a quantization of the gravitational field described in terms of the basis in the space of the fundamental spinor field. Thus the quantum theoretical description of the tetrad field or the metric field respectively appears as a consequence of a quantization concerning spinors assumed to be more fundamental in this approach.

\section{Preparation}

\subsection{Unified Quantum Field Theory of Spinors}

In the framework of relativistic quantum field theories elementary particles are described by irreducible representations of the
Lorentz group \cite{Wigner:1939,Weinberg:1995mt}. The simplest representation of the Lorentz group is given within a space of spinors representing particles of spin one half.
If we refrain from possible supersymmetric extensions, all the elementary matter fields of the standard model are described by spinor fields. All other kinds of fields like interaction fields with spin one can in principle be thought to be be composed from spinor fields. Therefore Heisenberg suggested a unified quantum field theory of a fundamental spinor field describing all matter fields and their interactions \cite{Heisenberg:1957,Heisenberg:1967,Heisenberg:1974du}.
The masses and interactions of particles in this theory are a consequence of a self interaction term of the elementary spinor field. In terms of Weyl spinors the postulated fundamental field equation reads

\begin{equation}
i\sigma^\mu \partial_\mu \psi \pm l^2 \sigma^\mu \psi \bar \psi \sigma_\mu \psi=0,
\label{fieldequation_Weyl}
\end{equation}
where $\psi$ denotes a Weyl spinor, $\bar \psi$ the adjoint Weyl Spinor and the $\sigma^\mu$ denote the Pauli matrices with the unit matrix $\sigma^0$ in two dimensions included. The quantity $l$ represents a fundamental constant of nature having the dimension of a length.
It is important to mention that gravity was omitted in Heisenberg's original setting of the theory which was founded on the postulate of symmetry with respect to the Lorentz group and the $SU(2)$ symmetry group with respect to the weak isospin. 
Vector bosons of spin one mediating the interactions within the standard model can be seen as states composed by a spinor state describing a particle and a spinor state describing an antiparticle.
This is in accordance with the representation $D(\frac{1}{2},0) \otimes D(0,\frac{1}{2})=D(\frac{1}{2},\frac{1}{2}) \oplus D(0,0)$ of the Lorentz group. In the original setting of Heisenberg the elementary spinor field was a dublett under the $SU(2)$ isospin symmetry group but not under the symmetry groups $SU(3)_{flavour}$ and $SU(3)_{colour}$ of the strong interaction, which are only approximate symmetries according to Heisenberg's theory.
Approaches to incorporate the symmetries of the strong interaction and supersymmetry to the theory can be found in \cite{Durr:1979fi,Durr:1982fk,Durr:1982nn}. But this topic is of no interest here, because we are just dealing with its relation to the structure of general relativity. Thus the spinor can be seen as a multiplett concerning any internal degree of freedom additionally to the spin structure.

\begin{figure}[ht]
\centering
\epsfig{figure=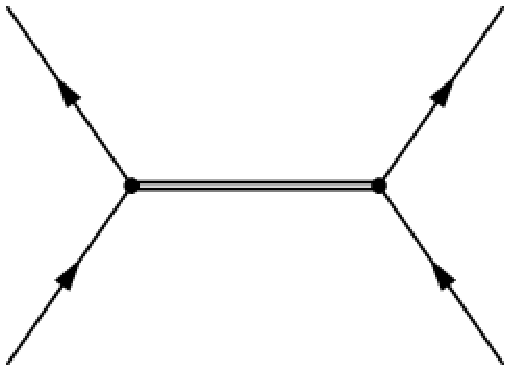,width=5cm}
\caption{\label{Feynman} Feynman graph of the self interacting fundamental spinor field: The horizontal double line represents a composed state from a spinor state describing a particle and an spinor state describing an antiparticle. At the vertices there coincide a state of a free particle and a state of a free antiparticle moving backwards in time.}
\end{figure}

The field has to be quantized according to the quantization procedure for fermionic fields postulating anticommutation relations for the field giving rise to the exclusion principle in the framework of quantum field theory. Thus one is led to the following anticommutation relations for the field
\begin{equation}
\left\{\psi^\alpha(x,t) ,\bar \psi^\beta(x^\prime,t) \right\}=\delta^3 \left(x-x^\prime \right) \delta^{\alpha \beta},\\
\label{quantization_spinorfield}
\end{equation}
where $\left\{A,B\right\} \equiv AB+BA$.
An interaction process where fermions exchange vector bosons, two electrons exchanging a photon for example, can be interpreted as a particle state interacting with an antiparticle state moving backwards in time and building a composed state before this state splits and there are two separated states representing free particles again (see figure ($\ref{Feynman}$)).

The question if the fundamental constant $l$ in ($\ref{fieldequation_Weyl}$) plays the role of a smallest length like the Planck length will be considered later if we will have introduced a Lagrangian for gravity. Since the self interaction term is the origin of the masses and the interactions described by the standard model, it has in each case to be connected with the electroweak scale.

\subsection{Spinors and the Space Time Metric}

We consider an abstract complex two dimensional vector space which elements are two component Weyl spinors. The complex conjugated spinor to a spinor $\varphi$ in this space shall be denoted with $\bar \varphi$. An arbitrary spinor $\varphi$ can be mapped to a vector in Minkowski space according to

\begin{equation}
k^\mu=\bar \varphi \sigma^\mu \varphi.
\label{spinortovector}
\end{equation}
Let us further assume the space to be endowed with a symplectic structure induced by a skew symmetric scalar product which shall be denoted by $[\cdot,\cdot]$. The property that $[\cdot,\cdot]$ is skew symmetric means that for two arbitrary spinors $\varphi$ and $\chi$ the relation $[\varphi,\chi]=-[\chi,\varphi]$ is valid.
The two form $[\cdot,\cdot]$ can be expressed by a matrix

\begin{equation}
\epsilon_{\alpha\beta}=\begin{pmatrix} 0 & 1 \\ -1 & 0 \end{pmatrix},
\label{symplectic_matrix}
\end{equation}
implying that the skew symmetric scalar product of two spinors $\varphi$ and $\chi$ looks as following

\begin{equation}
[\varphi,\chi]=\epsilon_{\alpha\beta} \varphi^\alpha \chi^\beta.
\label{symplectic_innerproduct}
\end{equation}
Thus $\epsilon_{\alpha\beta}$ can be used to rise and lower indices of spinors.
Mapping a spinor $\varphi^\alpha$ to the adjoint spinor $\varphi_\alpha$ with respect to the skew scalar product ($\ref{symplectic_innerproduct}$) means to map the contravariant vector in Minkowski space $k^\mu$ obtained  from ($\ref{spinortovector}$) to the corresponding covariant vector $k_\mu$.
The group $Sp(2,\mathcal{C})$ leaving the skew scalar product ($\ref{symplectic_innerproduct}$) invariant is isomorphic to the $SL(2,\mathbb{C})$ and thus to the homogeneous Lorentz group giving rise to a relation to Minkowski space time.
From an arbitrary basis within the spinor space consisting of two spinors according to the two dimensions of the space, call them $\varphi$ and $\chi$, a Minkowski space time tetrad can be constructed in the following way

\begin{eqnarray}
e_0^m=\frac{1}{2}(\bar \varphi \sigma^m \varphi+\bar \chi \sigma^m \chi),\quad e_1^m=\frac{1}{2}(\bar \varphi \sigma^m \chi+\bar \chi \sigma^m \varphi), \nonumber\\
e_2^m=\frac{1}{2}i(\bar \varphi \sigma^m \chi-\bar \chi \sigma^m \varphi),\quad e_3^m=\frac{1}{2}(\bar \varphi \sigma^m \varphi-\bar \chi \sigma^m \chi).
\label{tetrad}
\end{eqnarray}
According to the relation

\begin{equation}
g_{\mu\nu}=e_{\mu}^m e_{\nu m}
\label{metric}
\end{equation}
the tetrad ($\ref{tetrad}$) corresponds to a metric tensor $g_{\mu\nu}$
with signature $(+ ,- ,- ,- )$. In ($\ref{metric}$) we have used the fact that we can get $e_{\mu m}$ from $e_{\mu}^m$ by using the dual spinors of $\varphi$ and $\chi$ according to ($\ref{symplectic_innerproduct}$). Remark that latin indices denote flat indices and greek indices denote curved indices. If we make the following choice for the basis of spinors

\begin{equation}
\varphi=\begin{pmatrix} 1\\0 \end{pmatrix}\quad,\quad \chi=\begin{pmatrix} 0\\1 \end{pmatrix},
\end{equation}
($\ref{tetrad}$) and ($\ref{metric}$) yield the metric of flat Minkowski space time
$g_{\mu\nu}=(1,-1,-1,-1) \equiv \eta_{\mu\nu}$.
According to these considerations, there is a natural correspondence between a two dimensional spinor space endowed with a symplectic structure and Minkowski space time. A more elaborated treatment of these concerns can be found in \cite{Penrose:1985jw,Penrose:1986ca,Stewart:1990uf}.

\section{General relativity and the structure of space time from the connection of a fundamental Spinor field}

\subsection{Connection of the Spinor Field and Metric Structure}

According to Heisenberg we suggest that matter and its interactions are described by a fundamental spinor field. Heisenberg's theory in the original setting is formulated on a given Minkowski background. Approaches to incorporate gravity and to formulate the spinor field theory in a general relativistic setting can be found in \cite{Treder:1967,Durr:1982ie,Durr:1983uw}. But there the metric structure of space time is presupposed and the theory is just formulated on a given space time background.
A formulation of general relativity in terms of spinors is postulated in twistor theory \cite{Penrose:1960eq,Penrose:1977in}. There space time vectors themselves are a consequence of a postulated underlying spinor structure.
In this paper there is pursued another aim. An attempt is made to derive the properties of the gravitational field and thus the metric structure of space time from the properties of the abstract internal space of the fundamental spinor field. At the beginning we just assume that there exists a self interacting fundamental spinor field $\psi(x^\mu)$ on a four dimensional manifold representing space time before the introduction of a metric structure. The corresponding spinor space shall be endowed with a symplectic structure according to ($\ref{symplectic_matrix}$) and ($\ref{symplectic_innerproduct}$).
If one wants to compare two values of the spinor field at two different space time points, one has to define a spin connection. Such a spin connection, call it $A_{\mu\beta}^{\alpha}$, gives a prescription how to do this and it represents the property that it is possible to choose at every space time point another basis of spinors being equivalent to the definition of a nontrivial connection.
Since there can be chosen arbitrary coordinates, one is led to the connection group $GL(2,\mathbb{C})$. The $GL(2,\mathbb{C})$ has the $SL(2,\mathbb{C})$ and thus the Lorentz group as a subgroup. According to this connection one can define a covariant derivative $\nabla_\mu$ with respect to the spinor space

\begin{equation}
\nabla_\mu=\partial_\mu {\bf 1}+i A_{\mu\beta}^{\alpha}.
\label{covariant_derivative}
\end{equation}
Defining a spin connection $A_{\mu\beta}^{\alpha}$ due to ($\ref{covariant_derivative}$) is equivalent to the definition of two independent spinor fields, call them $\varphi$ and $\chi$, depending on the space time point, building a basis of spinors and being constant with respect to the covariant derivative ($\ref{covariant_derivative}$) which means that they are defined by the following relations

\begin{eqnarray}
\nabla_\mu \varphi^\alpha=\partial_\mu \varphi^\alpha+iA_{\mu\beta}^{\alpha} \varphi^\beta=0,\nonumber\\
\nabla_\mu \chi^\alpha=\partial_\mu \chi^\alpha+iA_{\mu\beta}^{\alpha} \chi^\beta=0.
\label{spinor_basis}
\end{eqnarray}
It makes no difference if we assume the connection $A_{\mu\beta}^{\alpha}$ or the basis of spinors consisting of $\varphi$ and $\chi$ to be more fundamental. Both representations contain the information how to compare values of the fundamental spinor field $\psi(x^\mu)$ at different space time points. Therefore it is also possible to define the connection by ($\ref{spinor_basis}$) in terms of the basis of spinors. If $\varphi$ and $\chi$ are given, then the connection $A_{\mu\beta}^{\alpha}$ fulfilling the relations ($\ref{spinor_basis}$) has the following shape

\begin{equation}
A_{\mu\beta}^{\alpha}=-i\frac{\partial_\mu \chi^\alpha \varphi_\beta-\partial_\mu \varphi^\alpha \chi_\beta}{\epsilon_{\gamma\delta}\varphi^{\gamma} \chi^{\delta}},
\label{connection}
\end{equation}
where we have used the symplectic expression $\epsilon_{\alpha\beta}$ introduced in ($\ref{symplectic_matrix}$) defining the relation between $\varphi^\alpha$ and $\varphi_\alpha$. In this sense the covariant derivative ($\ref{covariant_derivative}$) depends on $\varphi$ and $\chi$

\begin{equation}
\nabla_\mu=\partial_\mu {\bf 1}+i A_{\mu\beta}^{\alpha}(\varphi,\chi).
\label{covariant_derivative_basis}
\end{equation}
If we transport a spinor from one point to another, we are according to ($\ref{covariant_derivative}$) and ($\ref{spinor_basis}$) dealing with local basis transformations corresponding to a transition to a new value 
$\psi^\prime$ of the spinor field $\psi$ appearing in ($\ref{fieldequation_Weyl}$)

\begin{equation}
\psi=\psi_\varphi \varphi+\psi_\chi \chi \rightarrow \psi^\prime=\psi_\varphi \varphi^\prime+\psi_\chi \chi^\prime,
\label{fundamentalfield_basisrepresentation}
\end{equation}
which is equivalent to the old value with respect to the nontrivial connection $A_\mu^{\alpha\beta}$.

The tetrad defined above ($\ref{tetrad}$) leads for all spinors to a metric being proportional to the Minkowski metric.
In order to obtain general metrics we have to define an extended tetrad according to
\begin{eqnarray}
e^m_\mu=\frac{1}{2}
\left(\begin{matrix}
\bar \varphi \sigma^m \varphi+\bar \chi \sigma^m \chi\\
\bar \varphi \sigma^m \chi+\bar \chi \sigma^m \varphi\\
i\bar \varphi \sigma^m \chi-i\bar \chi \sigma^m \varphi\\
\bar \varphi \sigma^m \varphi-\bar \chi \sigma^m \chi
\end{matrix}\right)
+\frac{\bar \varphi \sigma^m \partial_\mu \chi-\bar \chi \sigma^m \partial_\mu \varphi}{\epsilon_{\alpha\beta}\varphi^\alpha \chi^\beta},
\label{general_tetrad}
\end{eqnarray}
where an additional term appears being equivalent to the connection expressed with one
Minkowski space index instead of two spin indices. For the case of a constant basis of spinor fields $\varphi$ and $\chi$ corresponding to a vanishing spinor connection the second term vanishes and the tetrad ($\ref{general_tetrad}$) reduces to ($\ref{tetrad}$)
leading to the special case of a flat Minkowski metric.

According to ($\ref{general_tetrad}$) and ($\ref{metric}$) from the basis of spinors we can construct a tetrad field $e_\mu^m(\varphi,\chi)$ and a metric field $g_{\mu\nu}(\varphi,\chi)$ respectively having signature $(+ ,- ,- ,- )$ because of the assumed symplectic structure and thus are led to a gravitational field.
Since the metric is constructed from the connection respectively the basis of spinor fields and since they shall constitute the metric structure of space time in our theory, the general covariance principle of general relativity is related to the arbitrary coordinate transformations within the spinor space. 
In accordance with usual general relativity our covariant derivative is defined in such a way that it
leaves the two spinor fields ($\ref{spinor_basis}$) constant defining an independent basis of spinors at each space time point.
Since the spinor connection leads to a tetrad due to ($\ref{general_tetrad}$) and to a metric with Minkowski signature due to ($\ref{metric}$), the fact that matter is composed by spinor fields could be seen as the reason why space time has a Lorentz structure and thus the covariant derivative ($\ref{covariant_derivative}$) also leaves the tetrad and the metric constant. This means that we can say that the spinor field is the origin of the metric structure.
\begin{figure}[ht]
\centering
\epsfig{figure=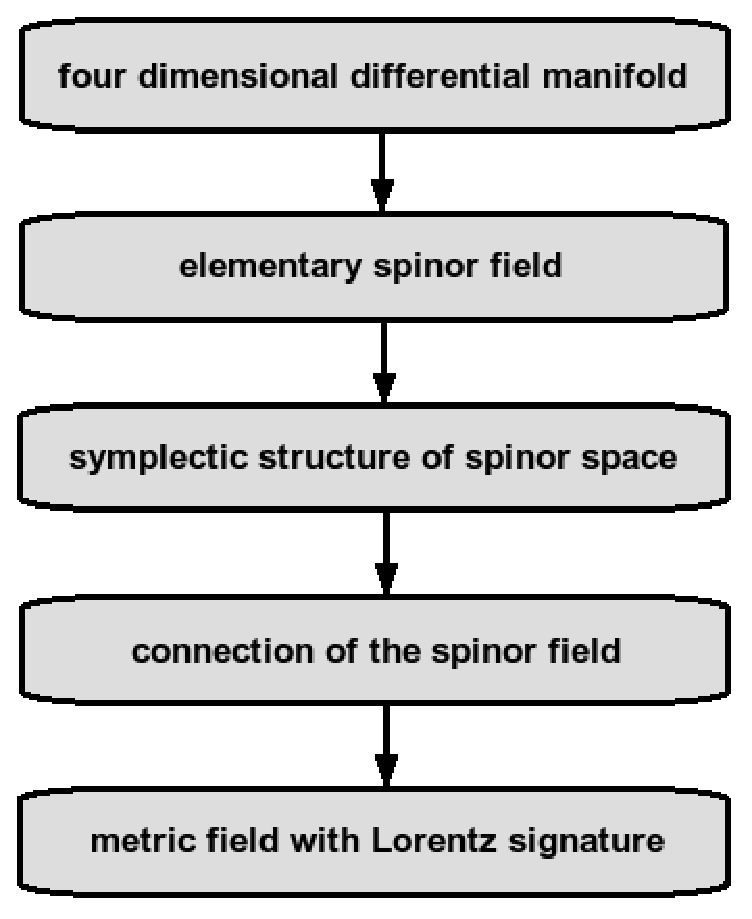,width=7cm}
\caption{\label{structure} Hierarchy of space time structure: At the beginning there is only assumed the structure of a differential manifold. On this manifold representing space time there is defined a spinor field and the metric properties of space time are derived from the properties of this abstract space of the spinor field.}
\end{figure}
This hierarchy concerning the origin of the space time structure, where the tetrad and the metric are derived quantities from an underlying spin structure of a fundamental matter field, is illustrated in figure ($\ref{structure}$).
The idea that a connection is more fundamental than the tetrad and the metric respectively is also hold in Ashtekar's new formulation of Hamiltonian gravity \cite{Ashtekar:1986yd,Ashtekar:1987gu,Ashtekar:1989ju} and the corresponding approach for the quantization of gravity. But there the connection is not associated with a fundamental matter field being described by spinors.

\subsection{Matter Action and Relation to the Gauge Description of Gravity}

In the usual gauge description of gravity (see \cite{Hehl:1976kj} for example) one begins with a Lagrangian of a matter field on Minkowski space time being invariant under global transformations of the Poincare group and postulates invariance under the corresponding local transformations. 
In \cite{Dehnen:1986mx} there has been suggested a spin gauge theory of gravity with gauge group $SU(2) \otimes U(1)$.
According to this approach one begins with a spinor field and postulates invariance under local transformations within the
space of Weyl spinors. This leads to a spin connection and from this one can define a field strength and thus one obtains a theory being equivalent to the linear approximation of general relativity. But this theory is not background independent. It not even supposes that there are perturbations of the flat Minkowski metric. According to the theory in \cite{Dehnen:1986mx} gravity is a phenomenon completely independent of the space time metric and is instead formulated on a Minkowski background given a priori. Besides it does not presuppose a unified theory of spinors and thus the interaction with gravity of interaction fields mediated by spin one particles is not included. 

In contrast to Heisenberg's theory \cite{Heisenberg:1957, Heisenberg:1967, Heisenberg:1974du} and the spin gauge description of gravity \cite{Dehnen:1986mx} we do not assume a Minkowski structure of space time and no metric field at the beginning.
It is just assumed that there exists a spinor field $\psi(x^\mu)$ defined on a four dimensional differentiable manifold representing space time from which all other fields are assumed to be composed. This leads to a background independent theory of gravity in a rigorous sense, because the gravitational field is a consequence of the spin connection of the fundamental matter field.
Since in our theory we do not assume any a priori metric structure, we cannot define an action and then use the gauge principle. We first have to define the general metric structure with the help of the properties of the spinor space and then we can build the action. Thus the fundamental action corresponding to a general relativistic setting of ($\ref{fieldequation_Weyl}$) has directly to be formulated with incorporation of the covariant derivative ($\ref{covariant_derivative}$) containing the connection of the spinor field and on a background $e_m^{\mu}(\varphi,\chi)$ respectively $g_{\mu\nu}(\varphi,\chi)$ derived from this connection respectively from the fields $\varphi$ and $\chi$ directly related to the connection ($\ref{covariant_derivative}$).
Thus we assume the action for the spinor field to look as following

\begin{eqnarray}
S_{m}&=&\int d^4 x \sqrt{-g(\varphi,\chi)}\left(i\bar \psi \sigma^m e^\mu_m(\varphi,\chi) \nabla_\mu \psi \right.\nonumber\\
&&\left.\quad\quad \pm \frac{l^2}{2} \bar \psi \sigma^\mu \psi \bar \psi \sigma_\mu \psi\right),
\label{matteraction}
\end{eqnarray}
where $g(\varphi,\chi)=\mbox{det}\left[g^{\mu\nu}(\varphi,\chi)\right]$ and thus $\sqrt{-g(\varphi,\chi)}=\mbox{det}\left[e^\mu_m\right(\varphi,\chi)]$. Note that according to ($\ref{covariant_derivative_basis}$) like $e_m^{\mu}(\varphi,\chi)$ and $g_{\mu\nu}(\varphi,\chi)$ the connection $A^{\alpha}_{\mu\beta}(\varphi,\chi)$ within the covariant derivative $\nabla_\mu$ depends also on $\varphi$ and $\chi$.
This action and the corresponding Lagrangian are invariant under $GL(2,\mathbb{C})$ transformations.

\subsection{Action of the Gravitational Field}

The action of the gravitational field has to be composed from the spin connection or the corresponding couple of spinor fields. In analogy to usual general relativity we can define a field strength tensor $F_{\mu\nu}^{\alpha\beta}(\varphi,\chi)$ (corresponding to the Riemann tensor) as the commutator of the covariant derivatives  

\begin{eqnarray}
F_{\mu\nu}^{\alpha\beta}(\varphi,\chi)&=&-i[\nabla_\mu,\nabla_\nu]
\nonumber\\
&=&\partial_\mu A_\nu^{\alpha\beta}(\varphi,\chi)-\partial_\nu A_\mu^{\alpha\beta}(\varphi,\chi)\nonumber\\
&&+iA_\mu^{\alpha\gamma}(\varphi,\chi)A_\nu^{\gamma\beta}(\varphi,\chi)-iA_\nu^{\alpha\gamma}(\varphi,\chi)A_\mu^{\gamma\beta}(\varphi,\chi),\nonumber\\
\label{fieldstrength}
\end{eqnarray}
where $[A,B] \equiv AB-BA$.
Remark that the connection $A_\mu^{\alpha\beta}$ appearing within the field strength tensor is defined by ($\ref{spinor_basis}$) and ($\ref{connection}$).
From this we can define a gravity action corresponding to the
action formulated within the spin gauge theory of gravity \cite{Dehnen:1986mx}. As already mentioned this theory is not background independent and it does not contain a self interacting spinor field assumed to be fundamental. However, we will use its dynamics in our approach which is shown to be equivalent to Einstein gravity with respect to the linearized approximation in \cite{Dehnen:1986mx}. This leads to the following gravity action

\begin{eqnarray}
S_{g}=\frac{1}{g}\int d^4 x \sqrt{-g(\varphi,\chi)}
g^{\mu\rho}(\varphi,\chi)g^{\nu\sigma}(\varphi,\chi)\nonumber\\
F_{\mu\nu}^{\alpha\beta}(\varphi,\chi)F_{\rho\sigma\alpha\beta}(\varphi,\chi),
\label{gravityaction}
\end{eqnarray}
where g is a fundamental constant describing the strength of gravity and being proportional to the usual gravitational constant G according to the relation

\begin{equation}
g=32\pi G.
\label{relationGg}
\end{equation}
As in the matter action there appear the spinor basis fields $\varphi$ and $\chi$ from the connection in the gravity action, which build a basis in the space of the fundamental matter field and define the tetrad field and the metric field.
Using ($\ref{fieldstrength}$) in ($\ref{gravityaction}$), the gravity action can be written more elaborately 

\begin{eqnarray}
S_{g}&=&\frac{1}{g}\int d^4 x \sqrt{-g(\varphi,\chi)}g^{\mu\rho}(\varphi,\chi)g^{\nu\sigma}(\varphi,\chi)\nonumber\\
&&\quad\quad\quad\left(2 \partial_{[\mu} A_{\nu]}^{\alpha\beta}(\varphi,\chi)\partial_{\rho}
A_{\sigma\alpha\beta}(\varphi,\chi)\right.\nonumber\\
&&\quad\quad\quad\left.
+4i \partial_{[\mu} A_{\nu]}^{\alpha\beta}(\varphi,\chi)
A_{\rho\alpha\gamma}(\varphi,\chi)A_{\sigma\gamma\beta}(\varphi,\chi)\right.\nonumber\\
&&\quad\quad\quad\left.
-2A_{[\mu}^{\alpha\gamma}(\varphi,\chi)A_{\nu]}^{\gamma\beta}(\varphi,\chi)
A_{\rho\alpha\delta}(\varphi,\chi)A_{\sigma\delta\beta}(\varphi,\chi)
\right).\nonumber\\ 
\label{gravity_action_connection}
\end{eqnarray}
The brackets $[\mu\nu]=\mu\nu-\nu\mu$ denote antisymmetrisation with respect to $\mu$ and $\nu$.
The dynamical behaviour of the tetrad field $e^\mu_m(\varphi,\chi)$ and the metric field $g_{\mu\nu}(\varphi,\chi)$ is also completely determined by ($\ref{gravityaction}$), since they are defined by $\varphi$ and $\chi$. If we want to consider the interaction of the gravitational field with matter described by the fundamental spinor field according to our approach, we have to include the corresponding Lagrangian ($\ref{matteraction}$). Thus the complete action appears as the sum of the matter action of the fundamental spinor field ($\ref{matteraction}$) and the gravity action ($\ref{gravityaction}$) related to the connection and the basis of spinor fields respectively. With this assumption the fundamental action of nature reads

\begin{eqnarray}
S&=&\int d^4 x \sqrt{-g(\varphi,\chi)}\nonumber\\
&&\left(\frac{1}{g}g^{\mu\rho}(\varphi,\chi)g^{\nu\sigma}(\varphi,\chi) F_{\mu\nu}^{\alpha\beta}(\varphi,\chi)F_{\rho\sigma\alpha\beta}(\varphi,\chi) \right.\nonumber\\
&&\left.\quad+ i\bar \psi \sigma^m e^\mu_m(\varphi,\chi) \nabla_\mu \psi \pm \frac{l^2}{2} \bar \psi \sigma^\mu \psi \bar \psi \sigma_\mu \psi\
\right).
\label{action}
\end{eqnarray}
It is obvious that besides the fundamental constants of special relativity and quantum theory, the speed of light c and Planck's constant h namely, which are set equal to one as usual, the constant l and the constant g defining a hierarchy between the action of the fundamental matter field $\psi$ and the action of the gravitational field represented by $\varphi$ and $\chi$ appear as the only fundamental constants in this theory. Thus the hierarchy between the electroweak and the Planck scale has to be a consequence of the relation between l and g, where l seems to play the role of a fundamental mass scale for particles and g is related to the usual gravitational constant G by ($\ref{relationGg}$). Because of ($\ref{relationGg}$) the Planck length representing a smallest length is directly related to g.
Thus l cannot be interpreted as a smallest length according to the unified spinor field theory of Heisenberg. Since the metric structure of space time is a consequence of the properties of the spinor space of matter, the fact that the action ($\ref{action}$) is invariant under arbitrary transformations of the $GL(2,\mathbb{C})$ reflects the general covariance of the matter and the gravity action.
Variation of ($\ref{action}$) with respect to $\bar \psi$ leads to the fundamental field equation for matter 

\begin{equation}
i\sigma^m e^\mu_m(\varphi,\chi) \nabla_\mu \psi \pm l^2 \sigma^\mu \psi \bar \psi \sigma_\mu \psi=0
\end{equation}
and variation with respect to $\bar \varphi$ and $\bar \chi$ leads to the fundamental equations for the gravitational field corresponding to Einstein's equation

\begin{eqnarray}
&&\left\{2 g^{\mu\rho}(\varphi,\chi)g^{\nu\sigma}(\varphi,\chi) \frac{\partial  F_{\mu\nu}^{\alpha\beta}(\varphi,\chi)}{\partial \bar \varphi}F_{\rho\sigma\alpha\beta}(\varphi,\chi)\right. \nonumber\\
&&\left.+2 \frac{\partial g^{\mu\rho}(\varphi,\chi)}{\partial \bar \varphi}g^{\nu\sigma}(\varphi,\chi)   F_{\mu\nu}^{\alpha\beta}(\varphi,\chi)F_{\rho\sigma\alpha\beta}(\varphi,\chi)
-e_{n \nu}(\varphi,\chi)\cdot\right. \nonumber\\
&&\left.\frac{\partial e^{n \nu}(\varphi,\chi)}{\partial \bar \varphi}
g^{\mu\rho}(\varphi,\chi)g^{\nu\sigma}(\varphi,\chi)
F_{\mu\nu}^{\alpha\beta}(\varphi,\chi)
F_{\rho\sigma\alpha\beta}(\varphi,\chi) \right\} \nonumber\\
&&=g\left\{-i\bar \psi \sigma^m \frac{\partial e^\mu_m(\varphi,\chi)}{\partial \bar \varphi} \nabla_\mu \psi
+e_{n \nu}(\varphi,\chi)\frac{\partial e^{n \nu}(\varphi,\chi)}{\partial \bar \varphi}\right.\nonumber\\
&&\left.\quad\quad\quad\quad \cdot\left[i\bar \psi \sigma^m e^\mu_m(\varphi,\chi) \nabla_\mu \psi \pm l^2 \bar \psi \sigma^\mu \psi \bar \psi \sigma_\mu \psi\right]\right\},\nonumber\\
\nonumber\\
&&\left\{2 g^{\mu\rho}(\varphi,\chi)g^{\nu\sigma}(\varphi,\chi) \frac{\partial  F_{\mu\nu}^{\alpha\beta}(\varphi,\chi)}{\partial \bar \chi}F_{\rho\sigma\alpha\beta}(\varphi,\chi)\right. \nonumber\\
&&\left.+2 \frac{\partial g^{\mu\rho}(\varphi,\chi)}{\partial \bar \chi}g^{\nu\sigma}(\varphi,\chi)  F_{\mu\nu}^{\alpha\beta}(\varphi,\chi)F_{\rho\sigma\alpha\beta}(\varphi,\chi)
-e_{n \nu}(\varphi,\chi)\cdot \right. \nonumber\\
&&\left.\frac{\partial e^{n \nu}(\varphi,\chi)}{\partial \bar \chi}
g^{\mu\rho}(\varphi,\chi)g^{\nu\sigma}(\varphi,\chi)
F_{\mu\nu}^{\alpha\beta}(\varphi,\chi)
F_{\rho\sigma\alpha\beta}(\varphi,\chi)\right\} \nonumber\\
&&=g\left\{-i\bar \psi \sigma^m \frac{\partial e^\mu_m(\varphi,\chi)}{\partial \bar \chi} \nabla_\mu \psi
+e_{n \nu}(\varphi,\chi)\frac{\partial e^{n \nu}(\varphi,\chi)}{\partial \bar \chi}\right.\nonumber\\
&&\left.\quad\quad\quad\quad \cdot\left[i\bar \psi \sigma^m e^\mu_m(\varphi,\chi) \nabla_\mu \psi \pm l^2 \bar \psi \sigma^\mu \psi \bar \psi \sigma_\mu \psi\right]\right\}.
\label{equations_gravity}
\end{eqnarray}
Here we have used that $\delta \sqrt{-g}=\delta \mbox{det} \left[e^\mu_m \right]=-\mbox{det} \left[e^\mu_m \right]e_{\nu n}\delta e^{\nu n}$ and
\begin{equation}
\delta_{\bar \varphi} e^\mu_m(\varphi,\chi)=\frac{\partial e^\mu_m(\varphi,\chi)}{\partial \bar \varphi}\delta \bar \varphi,\quad \delta_{\bar \chi} e^\mu_m(\varphi,\chi)=\frac{\partial e^\mu_m(\varphi,\chi)}{\partial \bar \chi}\delta \bar\chi,
\end{equation}
where $\delta_{\bar \varphi}$ denotes variation with respect to $\bar \varphi$ and $\delta_{\bar \chi}$ denotes variation with respect to $\bar \chi$. 
The terms arising from the matter action are written on the right hand sites of the equations (\ref{equations_gravity}). 
Thus the sum of their right hand sites, $\frac{\delta S_{matter}}{\delta \bar \varphi}+\frac{\delta S_{matter}}{\delta \bar \chi}$ namely, can be seen as the analogue to the energy momentum tensor appearing in Einstein's equations.
The dynamical behaviour of $e^\mu_m(\varphi,\chi)$ and $g^{\mu\nu}(\varphi,\chi)$ is indirectly determined by the above equations for $\varphi$ and $\chi$ leading to

\begin{eqnarray}
\partial_\lambda e^\mu_m(\varphi,\chi)=\frac{\partial e^\mu_m(\varphi,\chi)}{\partial \varphi}\partial_\lambda {\varphi}
+\frac{\partial e^\mu_m(\varphi,\chi)}{\partial \chi}\partial_\lambda {\chi},\\
\partial_\lambda g^{\mu\nu}(\varphi,\chi)=\frac{\partial g^{\mu\nu}(\varphi,\chi)}{\partial \varphi}\partial_\lambda {\varphi}
+\frac{\partial g^{\mu\nu}(\varphi,\chi)}{\partial \chi}\partial_\lambda {\chi}.
\end{eqnarray}

\subsection{Interpretation and Conceptual Issues}

From a conceptual or philosophical point of view background independence being connected to diffeomorphism invariance is the decisive property of general relativity. It is one of the central tasks of the search for a quantum theory of gravity to find a general relativistic setting of quantum field theories making allowance for this central principle. This topic is discussed elaborately in \cite{Rovelli:1999hz},\cite{Rovelli:2006yt} for example.
The gravity theory supposed here is background independent in an even more rigorous sense. In the usual setting of general relativity the metric structure of space time is not an absolute structure anymore like in special relativity. It becomes a dynamical entity itself. Since all matter fields live on space time, they all couple to gravity in the same manner. This is the origin of the equivalence principle. But conceptually they are separated from the gravitational field anyhow. There are in principle conceivable arbitrary types of fields interacting in arbitrary ways defined on space time which structure is described by general relativity. Space time connections are defined by the metric field representing gravity and defining its interaction with matter but being conceptually independent.
In the theory suggested here there is only the fundamental spinor field at the beginning. The connection of this matter field respectively the corresponding basis of spinor fields are the origin of the tetrad and thus the metric and not a metric field defined a priori. This is the reason why the principle of background independence seems to appear in an even more rigorous sense. The metric structure of space time reflects the properties of elementary matter and is not just dynamically related to it due to the field equations for the metric field.
This gives rise to a relationalistic attitude concerning the nature of space time, especially its metric structure.
Thus gravity can be seen as a gauge theory with respect to the fundamental spinor field of matter not presupposing a metric structure of space time but being the origin of it. As already mentioned this is in contrast to the usual gauge theoretic descriptions of gravity.

According to ($\ref{gravityaction}$) like matter the gravitational field itself can be described as a theory of spinors on a fundamental level in our theory. Therefore not only matter but also gravity itself has its origin in a spinor formulation.
This leads to a unification in a very radical sense which extends Heisenberg's theory which does not refer to gravity in the original setting. In the sense of ($\ref{general_tetrad}$) and ($\ref{metric}$) the tetrad description of the gravitational field although a consequence of an even more fundamental spinor description can be seen as more fundamental than the metric description anyhow.

Independent of these considerations the structure of space time as a four dimensional manifold has to be taken as a basic assumption in the approach presented in this paper. As already mentioned in the twistor approach of Penrose the structure of space time as a (3+1)-dimensional manifold is itself connected to an underlying spin structure. But there the question of a  unified description of matter fields is omitted.
In von Weizsaecker's reconstruction of physics \cite{Weizsaecker:1985, Lyre:2003tr} there is even derived the existence of a (3+1)-dimensional space time from the quantum theory of binary alternatives leading to spinors also. It has its origin in deep philosophical reflections about the meaning of quantum theory as a fundamental theory of nature not presupposing the structure of space time as a (3+1)-dimensional manifold but having it as a consequence.

\section{Approach for a Quantization of Gravity}

\subsection{Program for a Quantization of the fundamental Spinor Fields describing Gravity}

In the framework of the canonical quantization of gravity (see \cite{Kiefer:2004gr,Kiefer:2005uk,Giulini:2006xi} for example) there is performed a foliation of space time $\Sigma \times \mathcal{R}$ by choosing a spacelike hypersurface $\Sigma$ and thus separating a timelike direction $\mathcal{R}$. Then there is introduced an induced metric $h^{ab}$ on the three dimensional submanifold $\Sigma$, where a and b describe spatial coordinates. The definition of the time coordinate allows to define a canonical conjugated momentum $\pi^{ab}$ corresponding to $h^{ab}$ by referring to the usual Einstein Hilbert action expressed in terms of the new variables and thus one is led to a Hamiltonian $\mathcal{H}$ for gravity formulated in terms of $h^{ab}$ and $\pi^{ab}$. In the approach of quantum geometrodynamics based on these quantities there are postulated commutation relations between $h^{ab}$ and $\pi^{ab}$ and thus $h^{ab}$ and $\pi^{ab}$ become operators acting on quantum states $\Psi\left[h^{ab}(x)\right]$ depending on $h^{ab}(x)$.
After this quantization procedure there have to be implemented constraints to the states $\Psi\left[h^{ab}(x)\right]$  to get the space of real physical states. In the sense suggested by Dirac, the constraints are implemented as conditions on the states $\Psi\left[h^{ab}(x)\right]$. 
Since in our approach the spin connection of the fundamental matter field is assumed to be fundamental for gravity and the tetrad and the metric respectively are a consequence of the combination of the spinors representing a basis within the spinor space, the quantization of the tetrad field and the metric field have to be a consequence of the quantization of the more fundamental spinor connection structure connected to the spinor fields $\varphi$ and $\chi$.
In contrast to the quantization according to quantum geometrodynamics it makes no sense to consider an induced metric referring to the submanifold describing the spatial part of space time after a splitting of space time. The reason is that the fields $\varphi$ and $\chi$ determine the complete metric whereas the component of the metric with the positive sign refers to the chosen time direction. Therefore there can be specified a time direction $t$ without a splitting of the metric. Thus the quantization conditions of $\varphi$ and $\chi$ and the corresponding canonical momenta will imply quantum theoretical properties referring to the complete metric. The canonical conjugated momenta have to be defined according to ($\ref{gravityaction}$) in the usual way with respect to the chosen time coordinate

\begin{equation}
\Pi_\varphi=\frac{\delta S_{g}}{\delta \partial_t \varphi} \quad,\quad \Pi_\chi=\frac{\delta S_{g}}{\delta \partial_t \chi}.
\label{canonical_momenta}
\end{equation}
The choice of the time coordinate can be performed according to the usual formulation of Hamiltonian gravity because the fields $\varphi$ and $\chi$ representing the connection of the spinor field have a metric structure as a consequence with respect to which there can be performed a foliation into a spacelike hypersurface and a time direction.
Remark that the time coordinate has no absolute meaning in this context. It corresponds to this foliation of space time 
into Cauchy hypersurfaces. The covariance of general relativity is maintained because of the equal status of all possible foliations of this kind (see \cite{Kiefer:2004gr} for example).

There arises the question how the fields $\varphi$ and $\chi$ have to be quantized, with commutation or with anticommutation relations. If we remember that $\varphi$ and $\chi$ represent a spinor basis of the matter field $\psi$ in ($\ref{fieldequation_Weyl}$) implying that $\psi$ can be expressed as a linear combination of them according to ($\ref{fundamentalfield_basisrepresentation}$), one has to postulate anticommutation relations to maintain the validity of the anticommutation relations of the matter field ($\ref{quantization_spinorfield}$). Thus the quantization rules read

\begin{eqnarray}
\left\{\varphi^\alpha(x,t),\varphi^\beta(x^\prime,t) \right\}&=&\left\{\Pi_\varphi^\alpha(x,t),\Pi_\varphi^\beta(x^\prime,t) \right\}=0,\nonumber\\
\left\{\varphi^\alpha(x,t),\Pi_\varphi^\beta(x^\prime,t) \right\}&=&i\delta^{\alpha\beta}\delta^3(x-x^\prime),\nonumber\\ 
\left\{\chi^\alpha(x,t), \chi^\beta(x^\prime,t) \right\}&=&\left\{\Pi_\chi^\alpha(x,t),\Pi_\chi^\beta(x^\prime,t) \right\}=0,\nonumber\\
\left\{\chi^\alpha(x,t),\Pi_\chi^\beta(x^\prime,t) \right\}&=&i\delta^{\alpha\beta}\delta^3(x-x^\prime),\nonumber\\
\left\{\varphi^\alpha(x,t),\chi^\beta(x^\prime,t) \right\}&=&\left\{\varphi^\alpha(x,t),\Pi_\chi^\beta(x^\prime,t) \right\}=\nonumber\\
\left\{\Pi_\varphi^\alpha(x,t),\chi^\beta(x^\prime,t) \right\}&=&\left\{\Pi_\chi^\alpha(x,t),\Pi_\varphi^\beta(x^\prime,t) \right\}=0.\nonumber\\
\label{anticommutation_relation}
\end{eqnarray}
This implies quantum states $\Psi\left[\varphi(x),\chi(x)\right]$ depending on $\varphi$ and $\chi$ on which $\varphi$ and $\chi$ as well as the canonical conjugated momenta $\Pi_\varphi$ and $\Pi_\chi$ are acting on as operators.
The Hamiltonian constraints can be specified by defining a Hamiltonian for $\varphi$ and $\chi$ from ($\ref{gravityaction}$) 

\begin{equation}
\mathcal{H}(\varphi,\chi)=\Pi_\varphi \partial_t \varphi+\Pi_\chi \partial_t \chi-\mathcal{L}(\varphi,\chi), 
\label{Hamiltonian}
\end{equation} 
with the Lagrangian defined by ($\ref{gravityaction}$) according to $S_g=\int d^4 x \mathcal{L}$ and looking as following

\begin{eqnarray}
\mathcal{L}(\varphi,\chi)=
\frac{1}{g}\sqrt{-g(\varphi,\chi)} g^{\mu\rho}(\varphi,\chi)g^{\nu\sigma}(\varphi,\chi)\nonumber\\
F_{\mu\nu}^{\alpha\beta}(\varphi,\chi)F_{\rho\sigma\alpha\beta}(\varphi,\chi).
\label{Lagrangian}
\end{eqnarray}
From ($\ref{Hamiltonian}$) one can formulate the dynamical constraints (see \cite{Weinberg:1995mt} for example) according to the Heisenberg picture of the dynamics of quantum theory described by the commutator of the field operators with the Hamiltonian

\begin{eqnarray}
\partial_t \varphi=i\left[\mathcal{H},\varphi \right] \quad, \quad \partial_t \chi=i\left[\mathcal{H},\chi \right].
\label{dynamics_Heisenberg}
\end{eqnarray}
The dynamics of the tetrad field and the metric field being composed from $\varphi$ and $\chi$ is also determined by the 
relations ($\ref{dynamics_Heisenberg}$) and thus by the Hamiltonian ($\ref{Hamiltonian}$). This leads to the equations

\begin{eqnarray}
\partial_t e^\mu_m(\varphi,\chi)=i\left[\mathcal{H},e^\mu_m(\varphi,\chi)\right],\\
\partial_t g^{\mu\nu}(\varphi,\chi)=i\left[\mathcal{H},g^{\mu\nu}(\varphi,\chi)\right].
\end{eqnarray}
The quantum theoretical behaviour of the operators describing the tetrad field $e^\mu_m(\varphi,\chi)$ and the metric field $g_{\mu\nu}(\varphi,\chi)$ respectively have to be considered as derived from the quantization rules of $\varphi$ and $\chi$.

\subsection{Linearized Approximation}

The full gravity action expressed in terms of $\varphi$ and $\chi$ and the corresponding canonical momenta have a very complicated mathematical structure. Therefore there will be considered the case of a linear approximation, where the fields $\varphi$ and $\chi$ are assumed to be roughly free fields without self coupling.
In such a linear approximation of gravity one can assume the fundamental spinor fields $\chi$ and $\varphi$ to be roughly constant. In appropriate coordinates this means

\begin{equation}
\varphi \approx \begin{pmatrix} 1\\0 \end{pmatrix}=\mbox{const} \quad,\quad \chi \approx \begin{pmatrix} 0\\1 \end{pmatrix}=\mbox{const}.
\end{equation}
With this assumption one can also consider $g_{\mu\nu}(\varphi,\chi)=e_\mu^m(\varphi,\chi) e_{\nu m}(\varphi,\chi)$ to be roughly constant and thus in ($\ref{gravityaction}$) one can set $g_{\mu\nu}(\varphi,\chi) \approx \eta_{\mu\nu}$ and the Lagrangian in terms of the connection $A_{\mu}^{\alpha\beta}$ appearing in ($\ref{gravity_action_connection}$) corresponding to ($\ref{Lagrangian}$) reads in such a linear approximation

\begin{equation}
\mathcal{L}(\varphi,\chi)=\frac{2}{g} \partial_\mu A_\nu^{\alpha\beta}(\varphi,\chi)\partial^{[\mu}
A^{\nu ]}_{\alpha\beta}(\varphi,\chi).
\label{Lagrangian_gravity_linear}
\end{equation}
To calculate the canonical momenta, we have to use the explicit expression of the gravity
action in terms of $\varphi$ and $\chi$. We obtain this expression by using ($\ref{connection}$) in ($\ref{Lagrangian_gravity_linear}$). This leads to the following expression for the gravity Lagrangian in terms of
$\varphi$ and $\chi$

\begin{eqnarray}
&&\frac{1}{g}F_{\mu\nu}^{\alpha\beta}(\varphi,\chi)F^{\mu\nu}_{\alpha\beta}(\varphi,\chi)\nonumber\\
&=&\frac{2}{g}\left[\frac{\partial_\nu \chi^\alpha \partial^{[\nu} \chi_\alpha \partial_\mu \varphi^\beta \partial^{\mu]} \varphi_\beta
-\partial_\mu \chi^\beta \partial^{[\nu} \chi_\beta \partial_\nu \varphi^\alpha \partial^{\mu]} \varphi_\alpha}{\varphi_\epsilon \chi^\epsilon}\right. \nonumber\\
&-&\left.\frac{\partial_\mu (\varphi_\gamma \chi^\gamma)
\partial^{[\mu} (\varphi_\delta \chi^\delta)\partial_\nu \chi^\alpha \partial^{\nu]} \varphi_\alpha}{\left(\varphi_\epsilon \chi^\epsilon\right)^3}\right.
\nonumber\\
&-&\left.\frac{\partial^{[\mu}(\varphi_\gamma \chi^\gamma)\left(
\partial_\nu \chi^\alpha \partial_\mu \varphi^\beta \partial^{\nu]} \chi_\alpha \varphi_\beta 
+\partial_\nu \varphi^\alpha \partial_\mu \chi^\beta \partial^{\nu]} \varphi_\alpha \chi_\beta\right)}{(\varphi_\epsilon \chi^\epsilon)^3}\right.
\nonumber\\
&+&\left.\frac{\partial^{[\mu}(\varphi_\gamma \chi^\gamma)\left(
\partial_\nu \chi^\alpha \partial_\mu \varphi^\beta \partial^{\nu]} \chi_\alpha \chi_\beta
+\partial_\nu \varphi^\alpha \partial_\mu \chi^\beta \partial^{\nu]} \varphi_\alpha \varphi_\beta\right)}{(\varphi_\epsilon \chi^\epsilon)^3}\right].\nonumber\\
\label{gravity_action_spinor_linearized}
\end{eqnarray}
From this we can get the canonical momenta $\Pi_\varphi$ and $\Pi_\chi$ ($\ref{canonical_momenta}$)

\begin{eqnarray}
\Pi_\varphi^\alpha
&=&\frac{2}{g}\left[\frac{4 \theta^\alpha_\varphi(\varphi_\epsilon \chi^\epsilon)^2-2\omega_\varphi \chi^\alpha
-\partial_\mu (\varphi_\gamma \chi^\gamma)\partial^{[\mu} (\varphi_\delta \chi^\delta)\partial^{0]} \chi^\alpha}
{(\varphi_\epsilon \chi^\epsilon)^3}\right.\nonumber\\
&&\quad\quad\left.\frac{+\chi^\alpha(\theta_\varphi^\beta \chi_\beta+\theta_\chi^\beta \varphi_\beta-\theta_\varphi^\beta \varphi_\beta-\theta_\chi^\beta \chi_\beta)}
{(\varphi_\epsilon \chi^\epsilon)^3}\right.\nonumber\\
&&\quad\quad\left.\frac{+\lambda_\varphi(\chi^\alpha-\varphi^\alpha)+2\omega_\varphi^{\alpha\beta}(\varphi_\beta-\chi_\beta)}
{(\varphi_\epsilon \chi^\epsilon)^3}\right],\nonumber\\
\nonumber\\
\Pi_\chi^\alpha
&=&\frac{2}{g}\left[\frac{4 \theta^\alpha_\chi(\varphi_\epsilon \chi^\epsilon)^2-2\omega_\chi \varphi^\alpha
-\partial_\mu (\varphi_\gamma \chi^\gamma)\partial^{[\mu} (\varphi_\delta \chi^\delta)\partial^{0]} \varphi^\alpha}
{(\varphi_\epsilon \chi^\epsilon)^3}\right.\nonumber\\
&&\quad\quad\left.\frac{+\varphi^\alpha(\theta_\varphi^\beta \chi_\beta+\theta_\chi^\beta \varphi_\beta-\theta_\varphi^\beta \varphi_\beta-\theta_\chi^\beta \chi_\beta)}
{(\varphi_\epsilon \chi^\epsilon)^3}\right.\nonumber\\
&&\quad\quad\left.\frac{+\lambda_\chi(\varphi^\alpha-\chi^\alpha)+2\omega_\chi^{\alpha\beta}(\chi_\beta-\varphi_\beta)}
{(\varphi_\epsilon \chi^\epsilon)^3}\right].\nonumber\\
\label{canonical_momenta_linearized}
\end{eqnarray}
Using these canonical momenta in ($\ref{anticommutation_relation}$) leads to the quantization rules of the linear approximation.
The corresponding Hamiltonian can be obtained by inserting ($\ref{gravity_action_spinor_linearized}$) and ($\ref{canonical_momenta_linearized}$) in ($\ref{Hamiltonian}$) and reads

\begin{eqnarray}
\mathcal{H}&=
&\frac{2}{g}\left[\frac{4 \theta^\alpha_\varphi(\varphi_\epsilon \chi^\epsilon)^2-2\omega_\varphi \chi^\alpha
-\partial_\mu (\varphi_\gamma \chi^\gamma)\partial^{[\mu} (\varphi_\delta \chi^\delta)\partial^{0]} \chi^\alpha}
{(\varphi_\epsilon \chi^\epsilon)^3}\right.\nonumber\\
&&\quad\quad\left.\frac{+\chi^\alpha(\theta_\varphi^\beta \chi_\beta+\theta_\chi^\beta \varphi_\beta-\theta_\varphi^\beta \varphi_\beta-\theta_\chi^\beta \chi_\beta)}
{(\varphi_\epsilon \chi^\epsilon)^3}\right.\nonumber\\
&&\quad\quad\left.\frac{+\lambda_\varphi(\chi^\alpha-\varphi^\alpha)+2\omega_\varphi^{\alpha\beta}(\varphi_\beta-\chi_\beta)}
{(\varphi_\epsilon \chi^\epsilon)^3}\right] \partial_0 \varphi_\alpha\nonumber\\
&+&\frac{2}{g}\left[\frac{4 \theta^\alpha_\chi(\varphi_\epsilon \chi^\epsilon)^2-2\omega_\chi \varphi^\alpha
-\partial_\mu (\varphi_\gamma \chi^\gamma)\partial^{[\mu} (\varphi_\delta \chi^\delta)\partial^{0]} \varphi^\alpha}
{(\varphi_\epsilon \chi^\epsilon)^3}\right.\nonumber\\
&&\quad\quad\left.\frac{+\varphi^\alpha(\theta_\varphi^\beta \chi_\beta+\theta_\chi^\beta \varphi_\beta-\theta_\varphi^\beta \varphi_\beta-\theta_\chi^\beta \chi_\beta)}
{(\varphi_\epsilon \chi^\epsilon)^3}\right.\nonumber\\
&&\quad\quad\left.\frac{+\lambda_\chi(\varphi^\alpha-\chi^\alpha)+2\omega_\chi^{\alpha\beta}(\chi_\beta-\varphi_\beta)}
{(\varphi_\epsilon \chi^\epsilon)^3}\right] \partial_0 \chi_\alpha
\nonumber\\
&-&\frac{2}{g}\left[\frac{\partial_\nu \chi^\alpha \partial^{[\nu} \chi_\alpha \partial_\mu \varphi^\beta \partial^{\mu]} \varphi_\beta
-\partial_\mu \chi^\beta \partial^{[\nu} \chi_\beta \partial_\nu \varphi^\alpha \partial^{\mu]} \varphi_\alpha}{\varphi_\epsilon \chi^\epsilon}\right. \nonumber\\
&-&\left.\frac{\partial_\mu (\varphi_\gamma \chi^\gamma)
\partial^{[\mu} (\varphi_\delta \chi^\delta)\partial_\nu \chi^\alpha \partial^{\nu]} \varphi_\alpha}{\left(\varphi_\epsilon \chi^\epsilon\right)^3}\right.
\nonumber\\
&-&\left.\frac{\partial^{[\mu}(\varphi_\gamma \chi^\gamma)\left(
\partial_\nu \chi^\alpha \partial_\mu \varphi^\beta \partial^{\nu]} \chi_\alpha \varphi_\beta 
+\partial_\nu \varphi^\alpha \partial_\mu \chi^\beta \partial^{\nu]} \varphi_\alpha \chi_\beta\right)}{(\varphi_\epsilon \chi^\epsilon)^3}\right.
\nonumber\\
&+&\left.\frac{\partial^{[\mu}(\varphi_\gamma \chi^\gamma)\left(
\partial_\nu \chi^\alpha \partial_\mu \varphi^\beta \partial^{\nu]} \chi_\alpha \chi_\beta
+\partial_\nu \varphi^\alpha \partial_\mu \chi^\beta \partial^{\nu]} \varphi_\alpha \varphi_\beta\right)}{(\varphi_\epsilon \chi^\epsilon)^3}\right].\nonumber\\
\label{Hamiltonian_linearized}
\end{eqnarray}
In ($\ref{canonical_momenta_linearized}$) and ($\ref{Hamiltonian_linearized}$) we have introduced the following definitions

\begin{eqnarray}
\omega_\varphi^{\alpha\beta}&\equiv& \partial^{[\mu}(\varphi_\gamma \chi^\gamma)\partial_\mu
\chi^{\beta}\partial^{0]}\varphi^{\alpha},
\nonumber\\
\omega_\chi^{\alpha\beta}&\equiv& \partial^{[\mu}(\varphi_\gamma \chi^\gamma)\partial_\mu
\varphi^{\beta}\partial^{0]}\chi^{\alpha},
\nonumber\\
\omega_\varphi&\equiv& \partial^{[0}(\varphi_\gamma \chi^\gamma)\partial_\mu
\chi^{\alpha}\partial^{\mu]}\varphi_{\alpha},
\nonumber\\
\omega_\chi&\equiv& \partial^{[0}(\varphi_\gamma \chi^\gamma)\partial_\mu \varphi^{\alpha}\partial^{\mu]}\chi_{\alpha},
\nonumber\\
\theta_\varphi^{\alpha}&\equiv& \partial_\mu \chi^\beta \partial^{[\mu} \chi_\beta \partial^{0]} \varphi^{\alpha},
\nonumber\\
\theta_\chi^{\alpha}&\equiv& \partial_\mu \varphi^\beta \partial^{[\mu} \varphi_\beta \partial^{0]} \chi^{\alpha},
\nonumber\\
\lambda_\varphi&\equiv& \partial^{[0}(\varphi_\alpha \chi^{\alpha})\partial_\mu \chi^\beta \partial^{\mu]}\chi_\beta,
\nonumber\\
\lambda_\chi&\equiv& \partial^{[0}(\varphi_\alpha \chi^{\alpha})\partial_\mu \varphi^\beta
\partial^{\mu]}\varphi_\beta.
\nonumber\\
\end{eqnarray}
This quantization procedure gives also rise to (anti-)commutation relations for the tetrad field $\left[e^m_\mu(x,t),e^n_\nu(x^\prime,t)\right]\not= 0$ and the metric field $\left[g^{\mu\nu}(x,t)g^{\rho\sigma}(x^\prime,t)\right]\not= 0$ respectively and thus to a quantum state depending on the metric $\Psi \left[g^{\mu\nu}(x)\right]=\Psi \left[g^{\mu\nu}(\varphi(x),\chi(x))\right]$. This means that the quantum theoretical description of the gravitational field is related to the quantization of the fundamental spinor field.

\section{Summary}

We have suggested that the space time structure of general relativity could be the consequence of a connection of a fundamental
self interacting spinor field defined on a four dimensional differential manifold representing space time before the introduction of gravity and a corresponding metric structure. In such a description background independence and general covariance seem to become even more rigorous than in usual general relativity since the gravitational field representing the metric structure of space time is directly connected to the properties of the spin structure of a matter field assumed to be fundamental. In this sense one could assert that the gravitational field and thus the metric structure of space time are not as fundamental as matter fields but are a consequence of these fields. Thus a relationalistic view of space time, at least its metric structure, appears in a completely new way in this approach. As a consequence of the fact that the dynamics of the connection is expressed by two spinor fields assumed to be fundamental for gravity the dynamical behaviour of the metric field is derived from a more fundamental action referring to these spinor fields. The quantization of the gravitational field also occurs in a completely new way, because the quantum theoretical description of the gravitational field is derived from more fundamental canonical quantization rules of the couple of spinor fields related to the representation of the fundamental spinor field describing matter. Thus the quantization of gravity is directly connected to the quantization of the fundamental spinor field being at the origin of all other fields.
Altogether the presented theory seems to represent a very interesting approach to reconcile the decisive conceptual assertions of quantum field theory and general relativity.\\
\\
$Acknowledgement$:
\\ I would like to thank Benjamin Koch for very helpful discussions.

\end{document}